\documentclass[preprint,useAMS,usenatbib]{mn2e}
\usepackage{amsmath}
\usepackage{natbib}
\usepackage{graphicx}
\usepackage{subfigure}
\usepackage{amssymb}
\usepackage{amsfonts}
\usepackage{verbatim}
\usepackage{times}
\usepackage[usenames]{color}
\newcommand{\HI}{H{\sc i}}

\def\la{{\hbox{$\lesssim$}}}

\title [{\it Suzaku} observation and distance of G332.5-5.6] {{\it Suzaku} observation and distance of supernova remnant G332.5-5.6}

\author[Zhu, Tian \& Wu]{H. Zhu$^{1,2}$, W. W. Tian$^{1,3}$ and D. Wu$^{1,4}$\\
$^1$Key Laboratory of Optical Astronomy, National Astronomical Observatories, Chinese Academy of Sciences, Beijing, 100012, China; tww@bao.ac.cn\\
$^2$University of Chinese Academy of Sciences, 19A yuquan Road, Shijingshan District, Beijing, 100049, China\\
$^3$Department of Physics \& Astronomy, University of Calgary, Calgary, Alberta T2N 1N4, Canada\\
$^4$College of Information Science {\&} Technology, Beijing Normal University, Beijing, 100875, China}

\begin{document}

\maketitle

\begin{abstract}
We analyze the {\it Suzaku} XIS data of the central region of supernova remnant G332.5-5.6. The X-ray data are well described by a single non-equilibrium ionization thermal model, {\tt vnei}, with an absorbing hydrogen column density of 1.4$^{+0.4}_{-0.1}$ $\times$ 10$^{21}$ cm$^{-2}$. The plasma is characterized by an electron temperature of 0.49$^{+0.08}_{-0.06}$ keV with subsolar abundances for O (0.58$^{+0.06}_{-0.05}$ solar value) and Fe (0.72$^{+0.06}_{-0.05}$ solar value) and slightly overabundance for Mg (1.23$^{+0.14}_{-0.14}$ solar value). It seems that the central X-ray emission originates from projection effect or evaporation of residual clouds inside G332.5-5.6. We estimate a distance of 3.0 $\pm$ 0.8 kpc for G332.5-5.6 based on the extinction-distance relation. G332.5-5.6 has an age of 7 - 9 kyr.\\
\end{abstract}

\section{Introduction}

Supernova remnant (SNR) G332.5-5.6 was identified as an SNR candidate by \citet{dun97} based on the morphology at 4.85 GHz and the lack of 60 $\mu$m emission. \citet*{rey07} and \citet{stu07} both analyzed the 1.4 and 2.4 GHz radio continuum data from Australia Telescope Compact Array and obtained an average radio spectral index of $\alpha$ = -0.7 $\pm$ 0.2 and $\alpha$ = -0.6 $\pm$ 0.1 respectively. \citet{stu07} analyzed the optical spectra of G332.5-5.6 and gave average ratios of [{\sc Nii}]/H$\alpha$ = 2.42, [{\sc Sii}]/H$\alpha$ = 2.10 and [{\sc Sii}]6717/6731 = 1.23 which are all well within the typical values of SNRs. Due to these evidences, the authors confirmed G332.5-5.6 as an SNR.\\

SNR G332.5-5.6 has an angular size of about 30$'$ at radio wavebands with three main structures: two straight and nearly parallel outer filaments at northeast and southwest, an extended  radio emission region at center with small angular internal substructures (\citealt*{rey07}; \citealt{stu07}). G332.5-5.6 shows highly linear polarization of 0.3-0.5 at 2.4 GHz. \citet*{rey07} mentioned that the high polarization might be artificial, because the Q/U intensity images might have more small scale structures than the total intensity image and the lack of short baselines excludes more total intensity emission than Q/U intensity emission. {\it ROSAT} PSPC observation reveals strong X-ray emission from the center of the remnant, which has similar morphology with the central radio emission. Besides a small bright point-like source in the northeastern filament, the extended X-ray emission in the two outer filaments is weak (see Figure 7 of \citealt*{rey07} and Figure 17 of \citealt{stu07}). The central X-ray emission can be well fitted by the thermal Raymond-Smith model with a temperature of about 0.25 keV (\citealt{stu07}). Neither a radio nor a X-ray pulsar is found in the vicinity of SNR G332.5-5.6. It seems that the central X-ray emission is not from a nonthermal pulsar wind nebular (PWN) but from cloud evaporation or thermal conduction (\citealt*{rey07}).\\

H$\alpha$ filaments are detected mainly in the central region and match well with the radio structure (\citealt{stu07}). The spectra diagnosis of [{\sc Sii}]6717/6731 suggests an average electron density of about 240 cm$^{-3}$. The H$\beta$/H$\alpha$ flux ratio indicates a reddening of $E(B - V)$ = 0.27 towards G332.5-5.6. These facts indicate the blast at the center of SNR G332.5-5.6 is propagated into a cool and dense environment. Distance to the remnant is estimated in the range of 3 to 4 kpc based on {\HI} absorption, reddening and the surface brightness-diameter relation (\citealt*{rey07}, \citealt{stu07}).\\

Our knowledge about G332.5-5.6 is still poor. In this paper we present the {\it Suzaku} X-ray Imaging Spectrometer (XIS) observations of SNR G332.5-5.6 to determine its basic parameters, e.g. shock velocity, age, evolutionary state and elemental abundances. We also re-estimate its distance and discuss the origin of its central X-ray emission.\\

\section{Data and data reduction}

The central region of SNR G332.5-5.6 was observed by the XIS on board of {\it Suzaku}, starting on 2007 August 18 at 16:39:41 and ending on 2007 August 19 at 20:57:24 (observation ID 502066010, PI: Reynold S.). The pointing direction was (Ra, Dec) = (16h42m56.04s, -54d30m48.6s). Only the CCD camera XIS0, 1 and 3 were used during the observation with clocking mode of 8s exposure per frame and editing mode of 3 $\times$ 3. Each CCD has field of view of 17.8$'$ $\times$ 17.8$'$ and angular resolution of about 2$'$.\\

Data reduction is performed using {\tt HEAsoft} version 6.15. We reprocess and screen the data with the command {\tt Aepipeline} by applying the up-to-date calibration file and then obtain an effective exposure time of 70.15 ks, 70.15 ks and 70.16 ks for XIS0, XIS1 and XIS3 respectively. {\tt XSELECT} version 2.4c is employed to extract images and spectra. We used {\tt Xisnxbgen} to generate the XIS non-X-ray background (NXB) image and spectra based on the night Earth data. The {\tt rmf} and {\tt arf} files are created by {\tt Xisrmfgen} and {\tt Xissimarfgen} respectively. To fit the spectra, we use the software {\tt XSPEC} version 12.8.1.\\

\section{Results}

\subsection{X-ray Image}

\begin{figure}
\centering
\includegraphics[width=0.45\textwidth, angle=0]{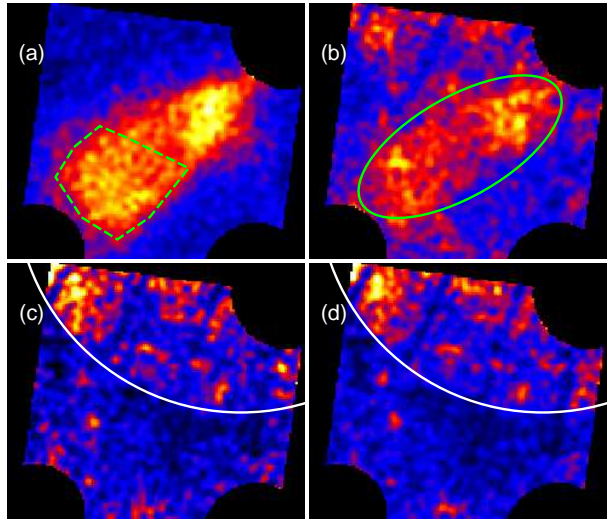}

\caption{The {\it Suzaku} XIS X-ray image of G332.5-5.6 in the energy band of 0.5-1.0 keV (a), 1.0-2.0 keV (b), 2.0-3.0 keV (c) and 2.0-5.0 keV (d). These images are smoothed with a Gaussian distribution of $\sigma$ = 24$"$. The green dashed polygons in (a) shows the on-SNR region. The white curves in (c) and (d) represent the regions affected by stray light. The ellipse in (b) occupies the same area as the ellipse in Figure \ref{fig2}.}
\label{fig1}
\end{figure}

\begin{figure}
\centering
\includegraphics[width=0.4\textwidth, angle=0]{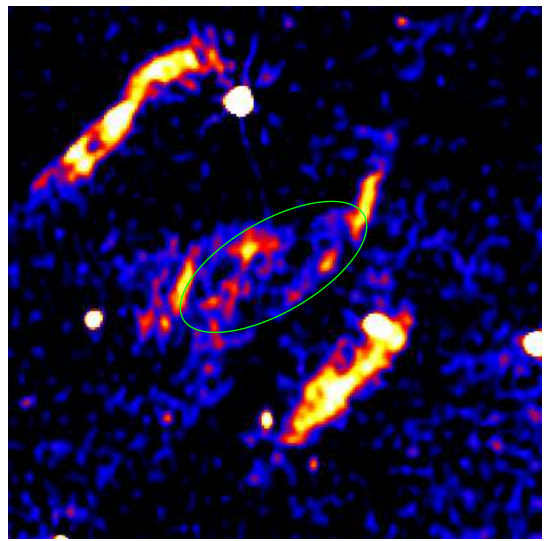}
\caption{The SUMSS 843 MHz image of G332.5-5.6 with an ellipse indicates the central X-ray emission region.}
\label{fig2}
\end{figure}

Figure \ref{fig1} shows the XIS0, 1 and 3 combined images of G332.5-5.6 in the energy band of 0.5-1.0 keV (a), 1.0-2.0 keV (b), 2.0-3.0 keV (c) and 2.0-5.0 keV (d). Figure \ref{fig2} displays the 843 MHz continuum image of G332.5-5.6 (data from the Sydney University Molonglo Sky Survey, \citealt{boc99}) with the X-ray bright region labeled by an ellipse. X-ray emission of G332.5-5.6 is distinct in the 0.5-1.0 keV and 1.0-2.0 keV bands with their morphology very similar to the extended central radio emission. There is nearly no emission above 2.0 keV. The central X-ray emission has two bright substructures centered on (Ra, Dec) = (16h43m15s, -54d33m17s) and (16h42m35s, -54d29m33s) respectively. The northwestern part of G332.5-5.6 marked by the white curves in Figures \ref{fig1}(c) and (d) is contaminated by stray light from a low mass X-ray binary 4U 1636-536 which is bright and about 50$'$ away from the G332.5-5.6. Further spectra analysis has avoided that region.\\


\subsection{X-ray spectra}

Figure \ref{fig1}(a) displays the region where we extract the spectra of the SNR (green dashed polygon, hereafter on-SNR). The on-SNR is outside the region contaminated by stray light. The background spectra are extracted from a region outside the SNR (hereafter off-SNR) and within the field of view of observation ID 502067010. The observation ID 502067010 points to (Ra, Dec) = (16h44m22s, -54d22m42s) and covers part of the northwest filament of G332.5-5.6. To increase the statistical quality, the front-side illuminated CCDs XIS0 and XIS3 spectra are merged. Then the SNR and the background spectra are binned to a minimum of 25 counts per bin to achieve a signal to noise ratio of 5.\\

\begin{figure}
\includegraphics[width=0.3\textwidth, angle=270]{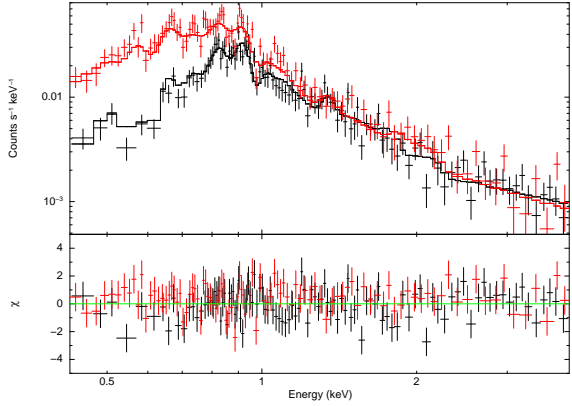}
\caption{The NXB-subtracted XIS0+XIS3 (black plus) and XIS1 (red plus) spectra of off-SNR. The solid black and red lines show the best-fit result.}
\label{fig3}
\end{figure}

Since we do not care for the real ingredient of off-SNR, the NXB-subtracted XIS0+XIS3 and XIS1 off-SNR spectra are fitted phenomenologically and simultaneously with the model {\tt phabs*(vnei+powerlaw)} within 0.4 keV to 4.0 keV. The fitting is showed in Figure \ref{fig3} with the reduced ${\chi}^{2}$ (dof) = 1.15 (198). We amplify the off-SNR model by area ratio between on-SNR and off-SNR and treat it as a fixed component in on-SNR spectra fitting.\\

\begin{figure}
\includegraphics[width=0.3\textwidth, angle=270]{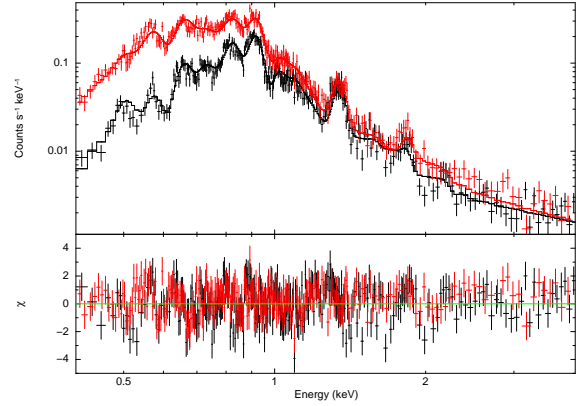}
\caption{The NXB-subtracted XIS0+XIS3 (black plus) and XIS1 (red plus) spectra of on-SNR. The solid black and red lines show the best-fit result using {\tt vnei} by changing the column density of hydrogen.}
\label{fig4}
\end{figure}

Figure \ref{fig4} shows the NXB-subtracted spectra of on-SNR. Line features (stronger than those in the off-SNR spectra) from {\sc O vii} (0.57 keV), {\sc O viii} (0.65 keV), {\sc O viii}/{Fe \sc xvii} (0.81 keV), {Ne \sc ix} (0.92 keV)and {Mg \sc xii} (1.35 keV) are clearly showed in the spectra, indicating thermal emissions from the SNR. We first try to fit the spectra with single collisional ionization equilibrium (CIE) models, {\tt vapec} and {\tt vmekal}. After several tests, we find that none of them can give acceptable fitting to the spectra (Both CIE models have reduced ${\chi}^{2}$ $\textgreater$ 1.43 and can not describe the spectra well in energy lower than 0.6 keV.). Thus we then turn to the non-equilibrium ionization (NEI) model, {\tt vnei}, which is characterized by a constant temperature T and single ionization timescale $\tau$ with flexible elemental abundances. The ionization timescale is defined as $\tau$ = n$_{e}$t, where n$_{e}$ is the post-shock electron density and t is the time since the passage of the shock. We set the {\tt vnei} version to 3.0 which uses the beta release of AtomDB 3.0 to calculate the spectra.\\

At the beginning of the {\tt vnei} fitting, both the plasma temperature T and ionization timescale $\tau$ are varied. The absorbing column density of hydrogen nuclei ${N}_{H}$ is fixed to $1.5 \times {10}^{21}$ cm$^{-3}$ based on the relation ${N}_{H}$ = $1.79 \times 10^{21}A_{V}$ (\citealt*{pre95}) with the extinction $A_{V}$ = 0.83 magnitude (\citealt{stu07}). The elemental abundance is set to the solar value. This fitting is poor with reduced ${\chi}^{2}$(dof) = 1.63 (536). Then we try to change the abundances of some of the elements in the spectra. Changing the abundance of oxygen can improve the fitting evident with a new ${\chi}^{2}$(dof) of 1.46 (535). The fitting can be further improved by changing the abundance of iron with ${\chi}^{2}$(dof) = 1.30 (534). Varying the abundances of magnesium improves the fitting with ${\chi}^{2}$(dof) of 1.28 (533). Changing the abundance of other elements, such as neon, gives little or no improvement to this fitting, therefore their abundances remains fixed to solar value. We also try to free the column density of hydrogen, then the ${\chi}^{2}$(dof) is 1.29 (532). The result is shown in Figure \ref{fig4} as solid lines. The best-fit parameters are listed in Table \ref{tab1}. The ionization timescale $\tau$ of on-SNR is small which can explain why CIE model could not fit the spectra as well. \\

\begin{table*}
\begin{minipage}[b]{1.0\textwidth}
\caption{Best-fit parameters for SNR G332.5-5.6 with model of {\tt vnei}, {\tt vapec} and {\tt vmekal}.$^a$}
\begin{center}
\begin{tabular}[t]{@{}p{2.0cm}p{1.5cm}p{1.5cm}p{1.5cm}p{1.5cm}p{1.8cm}p{2.5cm}p{2.5cm}@{}}
\hline
\hline
& & & & {\tt vnei} & &  & \\
\hline
${N}_{H} $(10$^{22} cm^{-2}$) & kT (KeV)  & O (solar)  & Mg (solar) & Fe (solar)  & $\tau$ (10$^{10} s\ cm^{-3}$)  &  Norm (10$^{-4}$ cm$^{-5}$)$^b$  & Reduced ${\chi}^{2}/dof$$^c$  \\

0.15 (fixed)   & 0.48$^{+0.04}_{-0.03}$ & 0.59$^{+0.04}_{-0.04}$ & 1.23$^{+0.14}_{-0.14}$ & 0.76$^{+0.02}_{-0.08}$ & 2.53$^{+0.61}_{-0.44}$ & 8.00$^{+1.04}_{-0.93}$ & 1.28/533 \\
0.14$^{+0.04}_{-0.01}$   & 0.49$^{+0.08}_{-0.06}$  & 0.58$^{+0.06}_{-0.05}$  & 1.23$^{+0.14}_{-0.14}$ & 0.72$^{+0.06}_{-0.05}$  & 2.53$^{+0.65}_{-0.54}$  & 7.64$^{+3.03}_{-2.04}$ & 1.29/532 \\
\hline
 & & & & {\tt vapec} & &  & \\
\hline
${N}_{H} $(10$^{22} cm^{-2}$) & kT (KeV)  & O (solar)  & Mg (solar) & Fe (solar) &  &  Norm (10$^{-2}$ cm$^{-5}$)  & Reduced ${\chi}^{2}/dof$ \\

0.31$^{+0.02}_{-0.03}$   & 0.18$^{+0.004}_{-0.004}$ & 0.54$^{+0.06}_{-0.05}$ & 2.12$^{+0.27}_{-0.27}$ & 2.30$^{+0.80}_{-0.58}$ & & 1.64$^{+0.25}_{-0.31}$  & 1.43/533 \\
\hline
& & & & {\tt vmekal} & &  &\\
\hline
${N}_{H} $(10$^{22} cm^{-2}$) & kT (KeV)  & O (solar)  & Mg (solar) & Fe (solar)  &   &  Norm (10$^{-2}$ cm$^{-5}$)  & Reduced ${\chi}^{2}/dof$  \\

0.27$^{+0.02}_{-0.02}$   & 0.18$^{+0.002}_{-0.002}$ & 0.54$^{+0.05}_{-0.04}$ & 2.31$^{+0.28}_{-0.26}$ & 1 (fixed) & & 1.27$^{+0.17}_{-0.16}$  & 1.53/535 \\
\hline
\end{tabular}
\end{center}
\begin{tabular}[t]{l}
$^a$ Errors are at the 90\% confidence level.\\
$^b$ $Norm = \frac{{{{10}^{ - 14}}}}{{4\pi {D^2}}}\int {{n_e}{n_H}dV}$, where D is the distance to G332.5-5.6 (cm), $n_e$ and $n_H$ (cm$^{-3}$) are the electron and hydrogen densities respectively.\\
$^c$ dof means the degrees of freedom.\\
\end{tabular}
\label{tab1}
\end{minipage}
\end{table*}

\section{Discussion}

\subsection{Distance to SNR G332.5-5.6}

Reliable distance to G332.5-5.6 is essential to estimate its basic parameters. \citet{stu07} mentioned G332.5-5.6 should be within the dust layer of the Galactic disc due to the rich shock-excited lines in its spectrum. Therefore it is feasible to build an extinction-distance relation to give restrictions to the distance of G332.5-5.6.\\

Red clump stars (RCs) are core helium-burning early $K$ giants which have obvious concentration region in a color-magnitude diagram (CMD, e.g. $J - K$ vs. $K$). Previous studies have shown that all RCs nearly have a uniform absolute magnitude and their infrared colors are insensitive to metal abundance. In view of this, RCs have been used as infrared standard candles (\citealt{lop02}, \citealt*{dur06}).\\

\begin{figure*}
\centerline{\includegraphics[width=0.33\textwidth, angle=0]{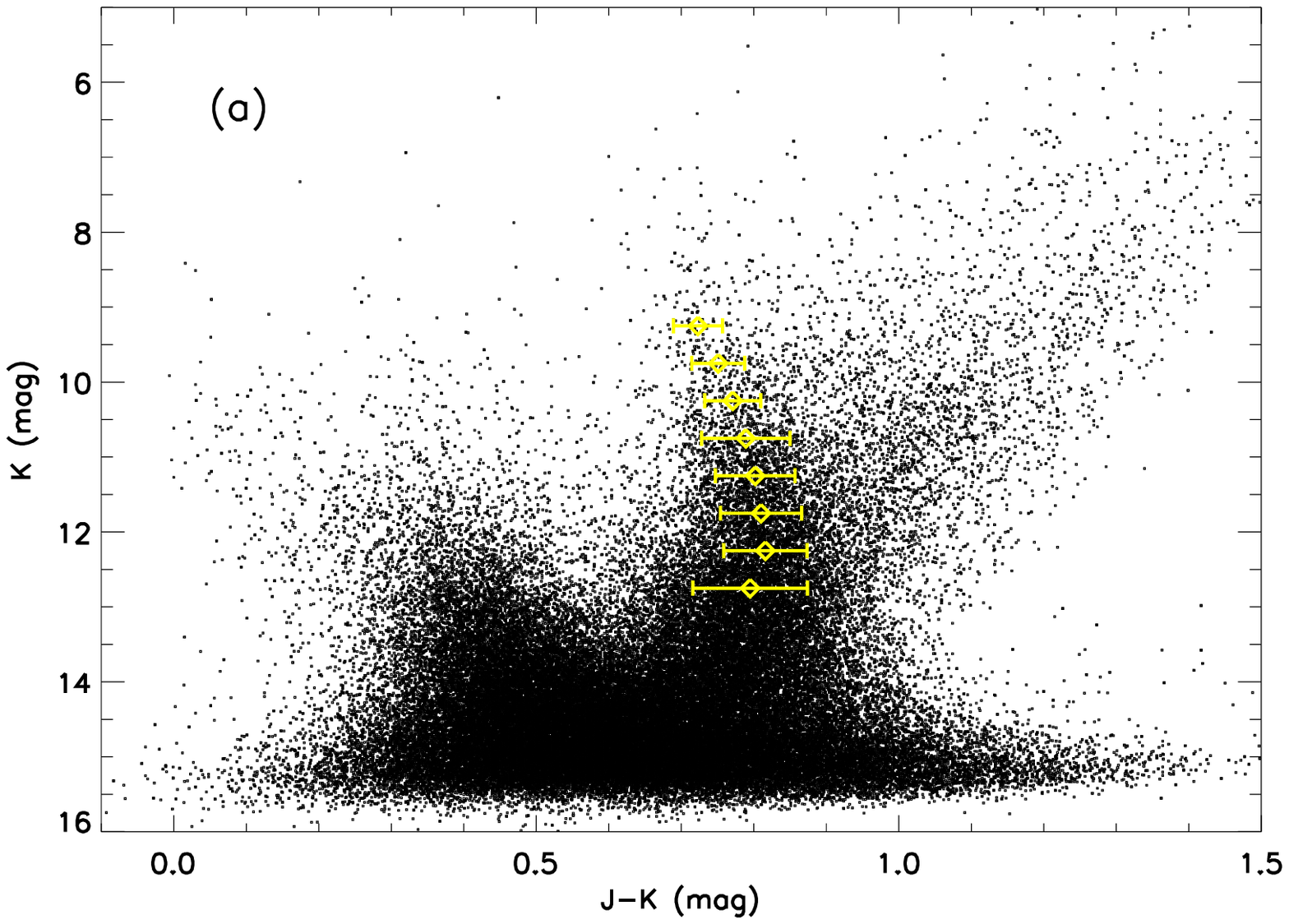}\includegraphics[width=0.33\textwidth, angle=0]{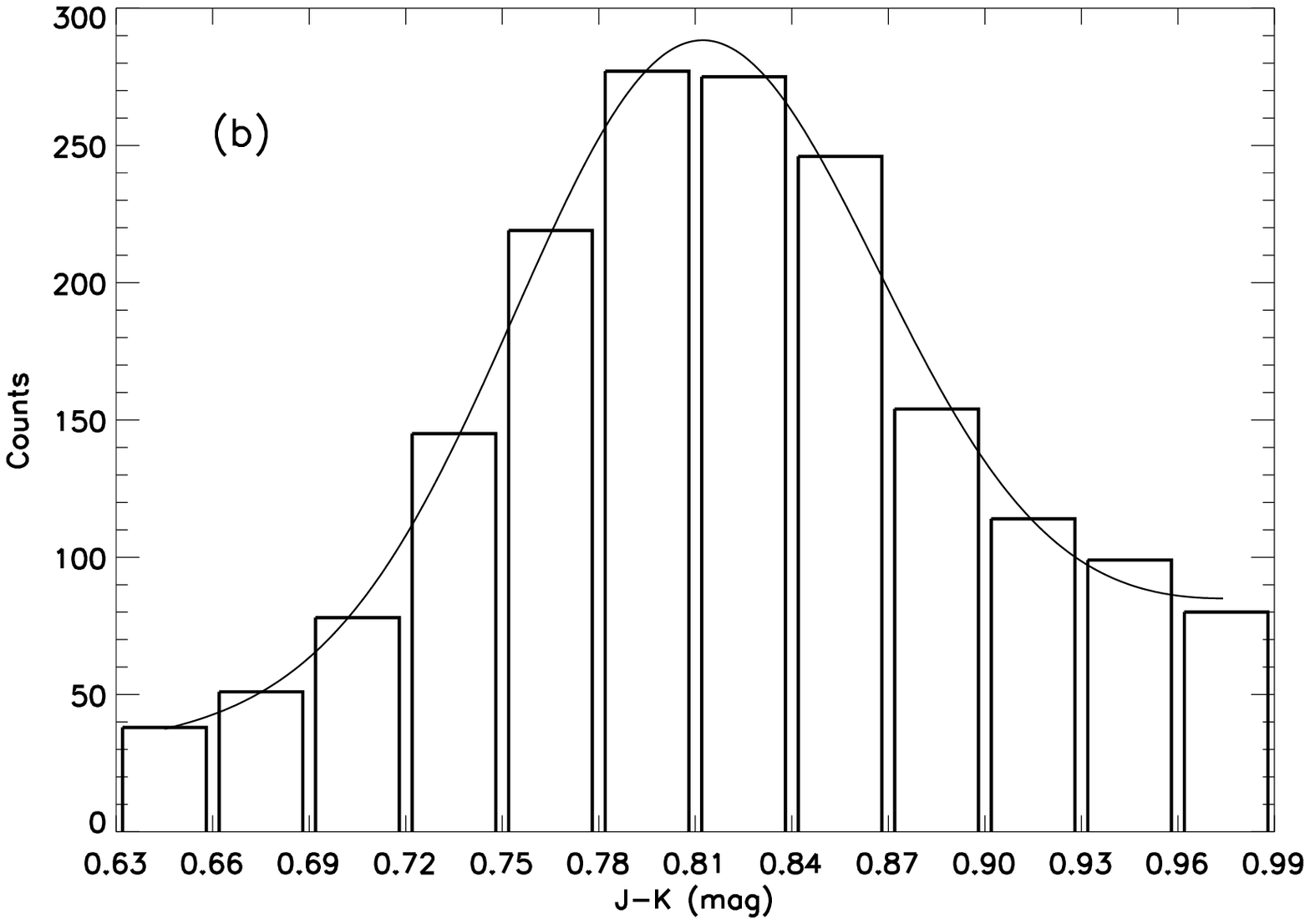}\includegraphics[width=0.33\textwidth, angle=0]{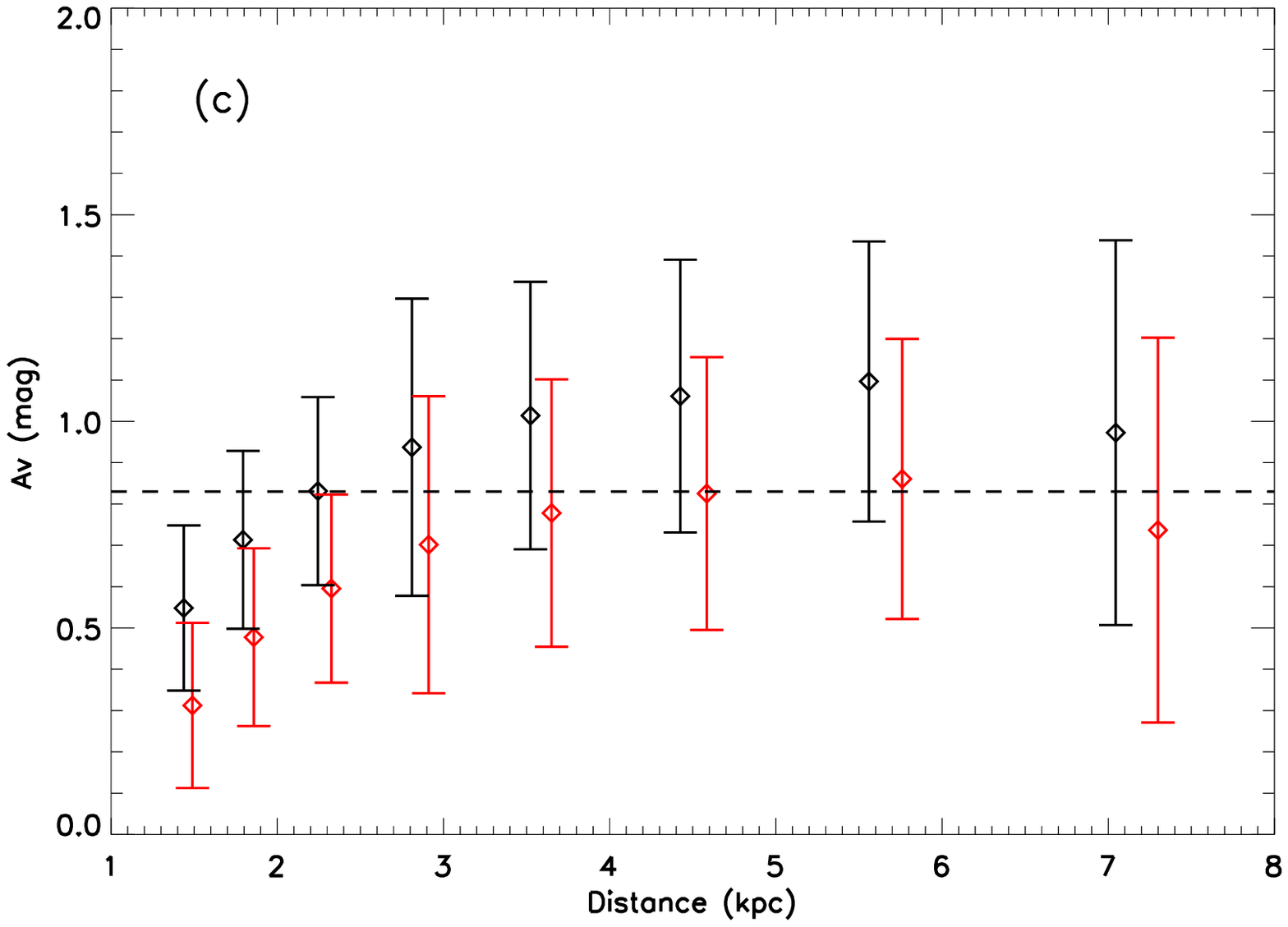}}
\caption{(a): CMD for 92020 stars within 0.5 deg$^2$ of G332.5-5.6. The yellow diamonds and yellow lines show the fitted location of the RCs peak and its extent with 1$\sigma$. (b): Histogram of the $J - K$ values of stars extracted from (a) in a 0.5 magnitude interval around $K$ = 11.75 mag. The black curve is the best-fit to this histogram using eq. (1). (c): The derived $V$ band extinction-distance relation along the direction towards G332.5-5.6 with overlapped $A_V$ value (dashed line) of 332.5-5.6. The red one is for $M_K$=-1.65 mag and $(J-k)_0$=0.67 mag. The black one is for $M_K$=-1.60 mag and $(J-k)_0$=0.63 mag.}
\label{fig5}
\end{figure*}

From the 2MASS All-Sky Point Source Catalog, we extract stars surrounding G332.5-5.6 in an area of 0.5 deg$^2$ with an interval of $\Delta${\tt l} = 1.0$^{\circ}$ and $\Delta${\tt b} = 0.5$^{\circ}$, a total number of 92020 stars. As seen in Figure \ref{fig5}(a), the region mainly occupied by RCs is distinct. We divide the $K$ band between 9 and 13 magnitude into 8 strips with a bandwidth of 0.5 magnitude. For each strip, the star counts are calculated in each $(J-k)$ bin of 0.3 magnitude. To decrease the influence of unrelated field stars, we only retain 5 bins on each side of the bin with maximum star counts. If the bin next to the maximum bin has similar counts with the maximum one, we will extend 5 bins to 6 bins. Figure \ref{fig5}(b) shows an example with the strip $K$ = 11.5-12.0 magnitude. In the example, the equation (\citealt*{dur06}):\\
\begin{equation}
y = {A_{count}}{(J - K)^\alpha } + {A_{RCs}}\exp \{ \frac{{ - {{[{{(J - K)}_{RCs}} - (J - K)]}^2}}}{{2{\sigma ^2}}}\},
\end{equation}
is used to fit the star counts distribution. The best-fit parameters are ${(J - K)}_{RCs}$ = 0.81 and $\sigma$ = 0.06. We assume the absolute $K$ band magnitude of RCs is $M_K$ = -1.60 to -1.65 mag and the intrinsic color of ${(J - K)}_{0}$ = 0.63 to 0.67 mag (values of \citealt*{alv00}, \citealt*{gro02}, \citealt{lan12} and \citealt{yaz13}). The extinction and distance follow (\citealt{guv10}, \citealt*{lop14}):\\
\begin{equation}
{A_V}(mag) = \frac{{A_K}}{0.1137} = \frac{{0.67[{{(J - K)}_{RCs}} - {{(J - K)}_0}]}}{{0.1137}},
\end{equation}
\begin{equation}
D(pc) = {10^{[({m_K} - {M_K} + 5 - {A_K})/5]}}.
\end{equation}

Figure \ref{fig5}(c) shows the extinction-distance relation along the line of sight towards G332.5-5.6. The dashed line represents the observed averaged extinction of G332.5-5.6. The extinction increases rapidly before 3.8 kpc and becomes nearly flat after that distance. Since the extinction is mainly caused by the dust, the extinction will grow quickly in the dust layer and grow slowly outside the dust layer. Therefore, the turnover can be used to indicate the boundary of dust layer. Because the observed rich shock-excited lines favor the remnant within the dust layer, 3.8 kpc should be an upper distance limit for the remnant. The lower distance limit is directly derived from the relation marked by the black points, i.e. 2.2 kpc. Considering all above, we suggest a distance of 3.0 $\pm$ 0.8 kpc for SNR G332.5-5.6. This distance is generally in agreement with previous distances (3 - 4 kpc, \citealt*{rey07}, \citealt{stu07}).\\

\subsection{The Origin of Central X-ray Emission}

Central X-ray emission could be formed by 1) nonthermal X-ray emission from a PWN, 2) thermal X-ray emission from the shocked ejecta, 3) enhanced central X-ray emission from thermal conduction effects which smooth the temperature and electron density distribution, 4) projection effect from X-ray emission of shocked interstellar medium and 5) evaporation of residual clouds inside the remnant.\\

\citet*{rey07} and \citet{stu07} had ruled out the possibility of nonthermal X-ray emission from a PWN because 1) the radio spectral index of the central region is not flatter than the index of two outer filaments 2) no pulsar has been discovered in the vicinity of G332.5-5.6 3) {\it ROSAT} data reveals thermal nature of the X-ray emission. By analyzing the {\it Suzaku} data, we find the central X-ray emission is thermal dominance with a subsolar, solar or slightly supersolar elemental abundances. This supports neither the PWN origin nor the shocked ejecta origin because the heavy elements of ejecta are usually overabundance. Thermal conduction, via Coloumb collisions between electrons and ions, can reduce the temperature and increase the density of the central region of the SNR. This has been suggested to explain the central filled X-ray morphology of mixed-morphology SNRs (\citealt{she99}, \citealt{sla02},). However this model usually produces a relatively flat surface brightness profile (see Figure 13 of \citealt{she99} and Figure 6 of \citealt{sla02}) which can not explain the steep X-ray profile of G332.5-5.6 (see Figure 18 of \citealt{stu07}). For projection effect and residual clouds evaporation, we can not distinguish them because both of them are harmonic with the observed data, i.e. element abundances and the obvious H${\alpha}$ emission in the X-ray emission region (Figure 3 of \citealt{stu07}).\\

\begin{table*}
\begin{minipage}[b]{1.0\textwidth}
\caption{Parameters of SNR G332.5-5.6}
\begin{center}
\begin{tabular}[t]{@{}p{3.5cm}p{2.5cm}p{2.5cm}p{2.5cm}p{2.5cm}@{}}
\hline
\hline
Parameters & projection Model  & \multicolumn{ 3}{c}{Evaporation Model}    \\
& & $C/\tau$ = 2.0 & $C/\tau$ = 2.5 & $C/\tau$ = 3.0 \\
\hline
Post-shock Temperature (keV)  & 0.49 & 0.43 & 0.54 & 0.68  \\
Shock radius (pc)  & 13.1 & 13.1 & 13.1  & 13.1 \\
Shock velocity (km s$^{-1}$)   & 650  & 600 & 680 & 760  \\
Age (kyr)  & 8 & 9 & 8 & 7  \\
\hline
\end{tabular}
\end{center}
\label{tab2}
\end{minipage}
\end{table*}

\subsection{The Nature of SNR G332.5-5.6}

{\it Suzaku} X-ray data of G332.5-5.6 can be described by plasma with temperature of 0.49 keV (5.7 $\times$ 10$^6$ K). This suggests the remnant is likely still in the adiabatic phase (\citealt{koo95}). \\

For the origin of projection effect, a shell SNR evolved in the uon-uniform and dense medium can form the mixed morphology (this model has an internal density profiles similar to those in the Sedov model, \citealt*{pet01}). Assuming that hydrogen and helium have a cosmic abundance ([He]/[H] = 0.1) and all of them are ionized, we get $n_e$ = 1.2$n_H$. The on-SNR region has an extent of about 7$'$. We assume a depth of 1.3$'$ (1/12 of the shock radius) and a filling factor, $f$, for the plasma. Under the distance of 3.0 kpc, the averaged post-shock electron density will be 0.53$/f$ cm$^{-3}$ or an averaged total particle density of 0.99$/f$ cm$^{-3}$. There exist the Rankine-Hugoniit jump conditions (\citealt*{rey08}),\\
\begin{equation}
\frac{{{\rho _2}}}{{{\rho _1}}} = \frac{{{u_1}}}{{{u_2}}} \approx \frac{{\gamma  + 1}}{{\gamma  - 1}} = 4,
\end{equation}
\begin{equation}
{T_2} \approx \frac{{2\gamma (\gamma  - 1)}}{{{{(\gamma  + 1)}^2}}}\frac{{\mu {m_p}}}{k}u_1^2 = \frac{3}{{16}}\frac{{\mu {m_p}}}{k}u_1^2.
\end{equation}
here, $\rho$, $u$ and $T$ mean density, velocity and temperature respectively. The subscripts 1 and 2 are for per-shock and post-shock. For monatomic plasma with solar abundances, $\gamma$ = 5/3 and $\mu$ = 0.6, the density of ambient medium is n$_0$ $\approx$ 0.25$/f$ cm$^{-3}$ and the shock velocity $V_s$ = $u_1$ $\approx$ 650 km s$^{-1}$. Taking the radius $R = \theta D = 13.1$ pc, we derive a characteristic age ${t_c} = \frac{2}{5}\frac{R}{{{V_s}}}$ $\approx$ 8 kyr.\\

For the evaporation model (\citealt*{whi91}), if we assume the density ratio between the cloud and the intercloud medium is large ($C \to \infty$) and the ratio of the evaporation timescale to the age of G332.5-5.6 is large ($\tau \to \infty$), the model will only depend on $C/\tau$ (\citealt{koo95}). The central X-ray morphology can be formed if $C/\tau$ $\ge$ 2. For $C/\tau$ = 2.0, 2.5 and 3.0, the post-shock temperature is equal to 0.43, 0.54 and 0.68 keV which indicate a shock velocity of about 600, 680 and 760 km s$^{-1}$. The characteristic age of G332.5 will be about 9, 8 and 7 kyr respectively. We list the derived parameters from both cases in Table \ref{tab2}.\\


\section{Summary}

We have analyzed the {\it Suzaku} XIS data of SNR G332.5-5.6. We find that the morphology of the central X-ray emission is resembles the radio emission. The X-ray data can not be fitted by a CIE model, neither vapec nor vmekal, but can be well described by a single NEI model, {\tt vnei}, with a temperature of 0.49$^{+0.08}_{-0.06}$ keV and subsolar abundances for O, Fe and slightly overabundance for Mg. The ionization timescale is 2.53$^{+0.65}_{-0.54}$ $\times$ 10$^{10}$ cm$^{-3}$ s which implies that the plasma is far from ionization equilibrium. This explains why CIE model can not fit the X-ray data well.\\

Using RCs as standard distance candles, we have derived a relationship between extinction and distance along the line of sight towards G332.5-5.6 and have estimated its distance of 3.0 $\pm$ 0.8 kpc. We discuss the origin of the central X-ray emission and find that the projection effect and evaporation of residual clouds likely cause the X-ray emission. Under both projection model and evaporation model, we derive the basic parameters of G332.5-5.6 (Table \ref{tab2}).\\

\section*{Acknowledgements}

We thank the referee for meaningful suggestions. HZ and WWT acknowledge supports from NSFC (211381001, Y211582001). This work is partly supported by China's Ministry of Science and Technology under the State Key Development Program for Basic Research (2012CB821800, 2013CB837901). We are grateful to Dr P. Zhou and S. Yao for their helps when preparing this paper. We also thank April D. Tian from UBC for proofreading this work.\\

\bibliographystyle{mn2e}


\end{document}